\documentclass[10pt]{iopart}

\usepackage{iopams}  

\usepackage{bm} % bold math
\usepackage{amssymb}
\usepackage{amsfonts}
\usepackage{mathrsfs}
\usepackage{comment}
\usepackage{graphicx}% Include figure files
\usepackage{dcolumn}% Align table columns on decimal point
\usepackage{bm}% bold math
\usepackage{psfig,epsfig}
\usepackage{fullpage}

\newcommand{\vecte}{{\mathbf{e}}}
\newcommand{\vectf}{{\mathbf{f}}}

\newcommand{\vectu}{{\mathbf{u}}}

\newcommand{\vectx}{{\mathbf{x}}}

\newcommand{\calO}{{\mathcal{O}}}

\newcommand{\bbN}{{\mathbb{N}}}

\newcommand{\bbR}{{\mathbb{R}}}

\newcommand{\re}{{Re}}

\newcommand{\sign}{{\mathrm{sign}\,}}

\newcommand{\x}{\langle \Psi'\Lambda\rangle}
\newcommand{\y}{\langle \Phi'\Lambda\rangle}
\newcommand{\PUV}{\langle \Psi UV\rangle}
\newcommand{\DFDP}{\langle\Phi'\Psi'\rangle}
\newcommand{\FP}{\langle\Phi\Psi\rangle}
\newcommand{\DFsq}{\langle{\Phi'}^2\rangle}
\newcommand{\DPsq}{\langle{\Psi'}^2\rangle}

\newcommand{\eps}{{\varepsilon}}

\newcommand{\E}{{\mathrm{e}}}
\newcommand{\D}{{\mathrm{d}}}

\newcommand{\eqref}[1]{(\ref{#1})}

\begin{document}

\title{Variational bounds on the energy dissipation rate 
        in body-forced shear flow}
\author{Nikola P Petrov\footnote{Also at the 
Michigan Center for Theoretical Physics, University of Michigan, 
Ann Arbor, MI 48109-1120, USA.}%
, Lu Lu 
and Charles R Doering\ddag
}
\address{Department of Mathematics, University of Michigan\\
Ann Arbor, MI 48109-1109, USA}

\eads{\mailto{npetrov@umich.edu}}
\eads{\mailto{lluz@umich.edu}}
\eads{\mailto{doering@umich.edu}}

\begin{abstract}
A new variational problem for upper bounds on the rate of energy
dissipation in body-forced shear flows is formulated by including a
balance parameter in the derivation from the Navier-Stokes equations.
The resulting min-max problem is investigated computationally, producing
new estimates that quantitatively improve previously obtained rigorous
bounds.  The results are compared with data from direct numerical
simulations.
\end{abstract}

\submitto{J.\ Turbulence}


\maketitle

%\tableofcontents

%%%%%%%%%%%%%%%%%%%%%%%%%%%%%%%%%%%%%%%%%%%%%%%%%%%%%%%%%%%%

\section{Introduction}  \label{sec:intro}

One of the outstanding open challenges for theoretical fluid mechanics in
the $21^\mathrm{st}$ century is to derive rigorous results for turbulence
directly from the fundamental equations of motion, the Navier-Stokes
equations, without imposing {\em ad hoc} assumptions 
or uncontrolled closures.
Exact results are extremely rare, but it is possible to derive rigorous
and physically meaningful limits on some of the fundamental physical
variables quantifying turbulent dynamics and transport.
The bulk rate of energy dissipation is one such quantity of particular
interest due to its production as a result of the turbulent cascade in the
high Reynolds number vanishing viscosity limit.
The derivation of mathematically rigorous bounds on the energy dissipation
rate, and hence also a variety important quantities such as turbulent drag
coefficients and heat and mass transport rates, has been a lively area of
research in recent decades.

Beginning in the early 1960s, L.N.~Howard and F.H.~Busse pioneered the
application of variational approaches for the derivation 
of rigorous---and physically relevant---bounds on the dissipation rate for
boundary-driven flows; see their reviews \cite{Howard72,Busse78}.
In the 1990s, P.~Constantin and the senior author of this paper introduced
the the so-called background flow method
\cite{DoeringC92,DoeringC94} based on an old idea by Hopf
\cite{Hopf41}.
The background method was soon improved by Nicodemus \etal
\cite{NicodemusGH97}  who introduced an additional variational `balance'
parameter, and by the late 1990s Kerswell \cite{Kerswell98} had shown that
the background method equipped with the balance parameter is dual to the
Howard-Busse variational approach.
Those theoretical techniques have been applied to many flows driven
by boundary conditions, including shear flows and a variety of thermal
convection problems 
\cite{ConstantinD95,ConstantinD96,%
DoeringW98,ConstantinD99,DoeringSW00,OteroWWD02,Otero04}.

Attention has recently turned as well to the derivation of
quantitative variational bounds on the energy dissipation rate for
body-forced flows.
In these systems, the bulk (space and time averaged) dissipation rate per
unit mass $\epsilon$ is proportional to the power required to maintain a
statistically steady turbulent state.
While body forces may be difficult to realize in experiments, they are
easily implemented computationally and are the standard method of driving
for direct numerical simulations (DNS) of turbulent flows.

Childress \etal \cite{ChildressKG01} applied a background-type method to
body-forced flows in a periodic domain, focusing on dissipation estimates
in terms of the magnitude of the applied force.
In dimensionless variables they bounded $\epsilon$ in units of
$(F^3\ell)^{1/2}$, where $F$ is the amplitude of the applied force per
unit mass and $\ell$ is the (lowest) length scale in the force.
The estimates were given in terms of the natural dimensionless control
parameter, the Grashof number, $Gr := F\ell^3/\nu^2$, where $\nu$
is
the kinematic viscosity.
In practice, $\epsilon$ is often measured in inviscid units of $U^3/\ell$
as a function of the Reynolds number $Re = U\ell/\nu$, where $U$ is a relevant
velocity scale---an emergent quantity when the force is specified {\it a
priori}.
In both cases the dissipation is bounded on one side by that of the
associated Stokes flow \cite{FoiasMT93}.
When bounds are expressed in terms of $Gr$, the Stokes limit is an upper
bound, whereas when the estimates are in terms of $Re$ it is the lower
limit.

Foias \cite{Foias97} was the first to derive an upper
bound on
\[
\beta := \frac{\epsilon \ell}{U^3} \
\]
in terms $Re$, but with an inappropriate prefactor dependence on the
aspect
ratio $\alpha = L/\ell$, where $L$ is the system volume, generally
an
independent variable from $\ell$ (see also
\cite{FoiasMRT01a,FoiasMRT01b}).
That analysis was recently refined by Foias and one of the authors of this
paper \cite{DoeringF02} to an upper estimate of the form
\[
\beta \leq c_1 + \frac{c_2}{Re} \ ,
\]
where the coefficients $c_1$ and $c_2$ are independent of $F, \ell, \nu$
and $\alpha$, depending only on the ``shape'' of the (square integrable)
body
force.
(This in consistent with much of the conventional wisdom about the cascade
in
homogeneous isotropic turbulence theory
\cite{Frisch95,DoeringG95,FoiasMRT01-book} as well as with wind tunnel
measurements \cite{Sreenivasan84} and DNS data \cite{Sreenivasan98}.)
Most recently, that approach was developed 
further by deriving a mini-max variational problem  on the time averaged
dissipation rate for a particular domain geometry \cite{DoeringES03}.
Moreover, the variational problem 
was solved exactly at high Reynolds numbers 
to produce estimates on the asymptotic behavior of
the energy dissipation as a function of $Re$ including 
the optimal prefactor.

In this paper we extend the results in \cite{DoeringES03} by introducing a
balance parameter $c$, the analog of the variational parameter introduced
by Nicodemus \etal \cite{NicodemusGH97, NicodemusGH98-I, NicodemusGH98-II}
for the background method.
This parameter controls a balance between the quantity being
bounded, the manifestly positive definite energy dissipation rate
proportional to the $L^2$ norm of the rate of strain tensor, and the
indefinite quantity derived from the power balance that is ultimately
being
extremized.
Specifically we consider the flow of a viscous incompressible fluid
bounded
by two parallel planes with free-slip boundary conditions at the walls and
periodic boundary conditions in the other two directions.
The flow is maintained by a time-independent body force in the direction
parallel to the walls.
First we derive the Euler-Lagrange equations in the case $c=0$ (where the
variational principle coincides with the one in \cite{DoeringES03}) and
solve them numerically at finite $Re$.
The full ($c>0$) Euler-Lagrange equations are quite complicated but they
can also be solved numerically by using Newton method with the $c=0$
solution
as an initial guess.

The rest of this paper is organized as follows.
In Section~\ref{sec:setup} we introduce the problem and its variational
formulation following \cite{DoeringES03}.
In Section~\ref{sec:bal-par} we present the augmented variational
problem and derive the variational equations, explaining how we go about
solving them.  
In Section~\ref{sec:num-results} we collect our numerical results, and in
Section~\ref{sec:concl} we summarize the results discussing the challenges
of this approach and future directions for research.

%%%%%%%%%%%%%%%%%%%%%%%%%%%%%%%%%%%%%%%%%%%%%%%%%%%%%%%%%%%%
%%%%%%%%%%%%%%%%%%%%%%%%%%%%%%%%%%%%%%%%%%%%%%%%%%%%%%%%%%%%

\section{Statement of the problem}  \label{sec:setup}

%%%%%%%%%%%%%%%%%%%%%%%%%%%%%%%%%%%%%%%%%%%%%%%%%%%%%%%%%%%%

\subsection{Notation}

Consider a viscous incompressible Newtonian fluid moving between two
parallel planes located at $y=0$ and $y=\ell$.
Denote $x$ the stream-wise direction and $z$ be the span-wise direction.
The velocity vector field satisfies free-slip boundary conditions at the
two
planes bounding the flow.  
We impose periodic boundary conditions in the other two directions.  
The motion of the fluid is induced by a steady body force $\vectf$ along
the
$x$ axis varying only in the $y$ direction.

The motion of the fluid is governed by Navier-Stokes equation
\begin{equation}  \label{eq:NS}
\frac{\partial \vectu}{\partial t}
        + (\vectu\cdot\nabla) \vectu + \nabla p
        = \frac{1}{Re} \Delta \vectu + \vectf
\end{equation}
and the incompressibility condition,
\begin{equation}  \label{eq:zero-div}
\nabla\cdot \vectu = 0 \ .
\end{equation}
Here $p(\vectx,t)$ is the pressure field, 
and $Re := \frac{U_\mathrm{rms} \ell}{\nu}$ is the Reynolds number, 
where $U_\mathrm{rms}$ is the root-mean square velocity of the fluid. 
The problem is non-dimensionalized by choosing 
the unit of length to be $\ell$ 
and the unit for time to be $\ell/U_\mathrm{rms}$.  
Let $\langle \cdot \rangle$ stand for the
space-time average.
With this choice of units the velocity of the fluid $\vectu(\vectx,t) =
(u,v,w)$ is space-time $L^2$-normalized to $1$:
\begin{equation}
\label{eq:normu}
\langle |\vectu|^2 \rangle =
\langle u^2 + v^2 + w^2 \rangle = 1  \ .
\end{equation}

Given $\epsilon$, is the space-time average energy dissipation rate in
physical units, the non-dimensional energy dissipation rate $\beta$  is
defined
\begin{equation}  \label{eq:beta0}
\beta := \frac{\ell \epsilon}{U_\mathrm{rms}^3} \ .
\end{equation}
The body force $\vectf$ in \eqref{eq:NS}
has the form
\[
\vectf(\vectx) = F \phi(y) \, \vecte_x \ ,
\]
where the dimensionless {\em shape} function $\phi:[0,1]\to\bbR$ has zero
mean and satisfies homogeneous Neumann boundary conditions,
and is $L^2$-normalized:
\[
\int_0^1 \phi(y) \, \D y = 0 \ , \quad
\phi'(0) = 0 = \phi'(1) \ , \quad
\int_0^1 \phi(y)^2 \, \D y = 1 \ .
\]
Now let $\Phi\in H^1([0,1])$ (where $H^p([0,1])$ is the space of functions
defined on $[0,1]$  with $L^2$-integrable $p^{th}$ derivatives) be the
{\em
potential} defined by
\[
\Phi' = -\phi \ , \quad \Phi(0) = 0 = \Phi(1) \ .
\]
(Note that we are free to impose homogeneous Dirichlet 
conditions on $\Phi$ at both boundaries 
due to the zero mean condition on $\phi$.)

The spatial domain is $(x,y,z)\in[0,L_x]\times[0,1]\times[0,L_z]$ where
$L_x$ and $L_z$ are the (non-dimensionalized) lengths in $x$ and $z$
directions.
Free-slip boundary conditions at the walls are realized by
\begin{equation}  \label{eq:BC0}
v = 0 \ , \quad
\frac{\partial u}{\partial y} = 0
= \frac{\partial w}{\partial y} \quad \mbox{ at } y=0 \ , 1 \ .
\end{equation}

%%%%%%%%%%%%%%%%%%%%%%%%%%%%%%%%%%%%%%%%%%%%%%%%%%%%%%%%%%%%

\subsection{Variational problem for the energy dissipation rate}

Here we follow \cite{DoeringES03} to derive the variational problem for
upper
bounds on the energy dissipation.
Multiplying Navier-Stokes equation \eqref{eq:NS} by $\vectu$, integrate
over the spatial domain, and average over time to obtain the energy
dissipation rate
\begin{equation}  \label{eq:eps_def}
\beta := \frac{1}{Re} \langle |\nabla\vectu|^2\rangle
        = \langle \vectf \cdot \vectu \rangle
        = F \langle \phi u \rangle
        = - F \langle \Phi' u \rangle  \ .
\end{equation}

To remove the explicit appearance of the amplitude $F$ of the body force,
multiply \eqref{eq:NS} by a vector field  of the form $\psi(y)\vecte_x$,
where the {\em multiplier function} $\psi\in H^2([0,1])$  satisfies
homogeneous Neumann boundary conditions $\psi'(0) = 0 = \psi'(1)$,
and is {\it not} orthogonal to the shape function $\phi$.
That is, $\langle \phi \psi \rangle \neq 0$.
We will also use the derivative of $\psi$
\[
\Psi\equiv \psi'\in H^1([0,1])
\]
which satisfies homogeneous Dirichlet boundary conditions $\Psi(0) = 0 =
\Psi(1)$ and is not orthogonal to the shape potential $\Phi$,
i.e., $\langle\Phi\Psi\rangle = \langle \phi \psi \rangle \neq 0$.
We will call $\Psi$ a {\em test function}.
Take the scalar product of \eqref{eq:NS} with $\psi(y)\vecte_x$, integrate
over the volume (integrating by parts by utilizing the boundary
conditions) and take the long-time average to see that
\begin{equation}  \label{eq:F1}
- \langle \Psi u v \rangle
        = \frac{1}{Re} \langle \Psi' u \rangle
        + F \langle \Phi \Psi \rangle \ .
\end{equation}
Express the amplitude $F$ of the body force  from \eqref{eq:F1} and insert
into the expression for the energy dissipation \eqref{eq:eps_def} to
obtain
\begin{equation}
\label{eq:beta1}
\beta =
 \frac{\langle \Phi' u\rangle \,
                \langle \Psi u v
                        + \frac{1}{Re}\Psi' u \rangle}
                {\langle \Phi \Psi \rangle}
\ .
\end{equation}

%%%%%%%%%%%%%%%%%%%%%%%%%%%%%%%%%%%%%%%%%%%%%%%%%%%%%%%%%%%%

\subsection{Mini-max upper bounds for $\beta$}

A variational bound on $\beta$ may be obtained by first maximizing the
right-hand side of \eqref{eq:beta1} over all unit-normalized
divergence-free
vector fields $\vectu$ that satisfy the boundary conditions
\eqref{eq:BC0},
and then minimizing over all choices of test functions  $\Psi\in
H^1([0,1])$
satisfying homogeneous  Dirichlet boundary conditions.
Then any solution of Navier-Stokes equation will have energy dissipation
rate $\beta$ bounded from above by
\begin{equation}   \label{eq:bound-DES}
\beta_b(Re) \equiv \min_{\Psi} \max_{\vectu}
        \frac{\langle \Phi' u\rangle \,
                \langle \Psi u v
                        + \frac{1}{Re}\Psi' u \rangle}
                {\langle \Phi \Psi \rangle}\ .
\end{equation}

In order to study the bound \eqref{eq:bound-DES} above, the authors of
\cite{DoeringES03} first evaluated (exactly)
\[
\beta_b(\infty) := \min_{\Psi} \max_{\vectu}
                \frac{\langle \Phi' u\rangle \,
                \langle \Psi u v \rangle}
                {\langle \Phi \Psi \rangle}\ ,
\]
and then used this result to analyze the behavior
of $\beta_b(Re)$ for finite $Re$.
Since we are going to generalize that approach,
we briefly recall the analysis:

The evaluation began with the proof that
\begin{equation}  \label{eq:re-inf}
\max_{\vectu} \, \langle\Phi' u\rangle \langle \Psi u v \rangle
= \frac1{\sqrt{27}} \, \sup_{y\in[0,1]} |\Psi(y)| \ .
\end{equation}
This was accomplished by showing that the right-hand side
of \eqref{eq:re-inf} is an upper bound
for $\langle\Phi' u\rangle \langle \Psi u v \rangle$
for any $\vectu$ in the class of vector field considered,
and then explicitly constructing a sequence of unit-normalized
divergence-free vector fields
$\vectu^{(k)}=(u^{(k)},v^{(k)},w^{(k)})$
satisfying the boundary conditions \eqref{eq:BC0}
such that $\vectu^{(k)}$ saturate this bound
in the limit $k\to\infty$, i.e.,
\[
\lim_{k\to\infty}
\langle\Phi' u^{(k)}\rangle \langle \Psi u^{(k)} v^{(k)} \rangle
=
\frac1{\sqrt{27}} \, \sup_{y\in[0,1]} |\Psi(y)| \ .
\]
The precise form of $\vectu^{(k)}$ is
\begin{eqnarray}  \label{eq:u-k}
u^{(k)}(y,z) &=& g_k(y) \sqrt{2} \sin{kz}
                - \frac{1}{\sqrt{3}} \Phi'(y) \nonumber \\
v^{(k)}(y,z) &=& g_k(y) \sqrt{2} \sin{kz} \\
w^{(k)}(y,z) &=& \frac{1}{k} g'_k(y) \sqrt{2} \cos{kz} \ , \nonumber
\end{eqnarray}
where the sequence $g_k$ consists of smooth functions
approximating as $k\to\infty$ a Dirac $\delta$ function with support
centered at the points where the function $\Psi\in H^1([0,1])$
reaches an extremum, and normalized as
\[
\left\langle g_k^2 + \frac{1}{2k^2} {g'_k}^2 \right\rangle
= \frac{1}{3} \ .
\]
Note that the function $\Psi\in H^1([0,1])$ is continuous and hence it
reaches its extremum in $[0,1]$.
Moreover, since $\Psi(0)=0=\Psi(1)$ and at the same time $\Psi$ is not
identically zero, a point where $\Psi$ reaches an extremum must be
in the open interval $(0,1)$.

Following \eqref{eq:re-inf}, it was proved that if $\Phi\in H^1([0,1])$
changes sign only finitely many times,  then
\[
\beta_b(\infty) =
\frac{1}{\sqrt{27}} \, \min_\Psi \sup_{y\in[0,1]}
        \frac{|\Psi(y)|}{\langle\Phi\Psi\rangle} =
\frac{1}{\sqrt{27}} \frac{1}{\langle|\Phi|\rangle} \ ,
\]
which is achieved for the choice of test function $\Psi = \sign \Phi$.
While $\sign \Phi$ is not in $H^1([0,1])$, it can be approximated
arbitrarily closely (in the sense of pointwise convergence) by a sequence
of
functions in $H^1([0,1])$.

In \cite{DoeringES03}, the authors considered test functions $\Psi_\delta$
which are ``linearly mollified''  approximations of $\sign\Phi$, i.e.,
continuous piecewise linear functions approximating $\sign \Phi$ by
replacing the jumps of $\sign \Phi$ by lines of slope $\pm\frac1\delta$
connecting the values $-1$ and $1$ (see Figure~1 in \cite{DoeringES03}).
Finally, for finite $Re$, it was shown in \cite{DoeringES03} that by
choosing $\delta\sim\calO(\re^{-1/2})$, the dissipation rate for $\Phi\in
H^1([0,1])$ behaves for large $Re$ as
\[
\beta_b(Re) \leq \beta_b(\infty) + \calO(Re^{-3/4}) \ .
\]
If $\Phi$ is smooth (i.e., $\Phi$ has a bounded derivative and so
behaves linearly around its zeroes), then by taking
$\delta\sim\calO(Re^{-2/5})$ it was shown as well that
\[
\beta_b(Re) \leq \beta_b(\infty) + \calO(Re^{-4/5}) \ .
\]

%%%%%%%%%%%%%%%%%%%%%%%%%%%%%%%%%%%%%%%%%%%%%%%%%%%%%%%%%%%%
%%%%%%%%%%%%%%%%%%%%%%%%%%%%%%%%%%%%%%%%%%%%%%%%%%%%%%%%%%%%

\section{Improved variational principle}
\label{sec:bal-par}

%%%%%%%%%%%%%%%%%%%%%%%%%%%%%%%%%%%%%%%%%%%%%%%%%%%%%%%%%%%%

\subsection{Introducing the balance parameter}

%We obtain bounds on the energy dissipation,
%we do the following:

Let $c\in[0,\infty)$ be arbitrary.
Multiply \eqref{eq:beta1} by $1+c$
and add it to
$\beta = \frac{1}{Re} \langle|\nabla\vectu|^2\rangle$
multiplied by $-c$.
The result is
\begin{equation}  \label{eq:beta_c}
\beta
=
(1+c) \frac{\langle \Phi' u\rangle \,
        \langle \Psi u v
        + \frac{1}{Re}\Psi' u \rangle}{\langle \Phi \Psi \rangle}
- \frac{c}{Re} \langle|\nabla\vectu|^2\rangle
\ .
\end{equation}
Now we will obtain bounds on the energy dissipation by applying a mini-max
procedure to the functional in the right-hand side above.

The parameter $c$ provides more constraint on the variational procedure
than
the case considered in \cite{DoeringES03}.
The space-time average of $|\nabla\vectu|^2$ is multiplied by $-c<0$ so
that
for a velocity field with a large gradient (like the one of the form
\eqref{eq:u-k}  when $g_k$ tends to a Dirac $\delta$ function), the
right-hand side of \eqref{eq:beta_c} will become smaller.

While performing the maximization procedure we have to incorporate two
explicit constraints on the velocity vector fields: the unit-norm
condition
\eqref{eq:normu}  and incompressibility \eqref{eq:zero-div}.
The former one is easy to implement by adding a term with Lagrange
multiplier $\lambda$  which is a number (i.e., does not depend on $\vectx$
and $t$).   Incompressibility, however, requires introducing
a Lagrange multiplier (a ``pressure'') that is a pointwise function which
makes the variational problem very difficult to analyze.
So instead we will restrict the class of velocity fields $\vectu$
over which we maximize to fields that are automatically divergence-free.

The functional incorporating the normalization constraint is
\begin{equation}  \label{eq:L}
L[\vectu]
:=
(1+c) \frac{\langle \Phi' u\rangle \,
        \langle \Psi u v
        + \frac{1}{Re}\Psi' u \rangle}{\langle \Phi \Psi \rangle}
- \frac{c}{Re} \langle|\nabla\vectu|^2\rangle
+ \frac{\lambda}{2} \, \langle|\vectu|^2-1\rangle \ .
\end{equation}
The  class of velocity fields $\vectu$ we will consider is a
generalization
of \eqref{eq:u-k}:
\begin{eqnarray}   \label{eq:ansatz}
u(y,z) &=& U(y) \, \sqrt{2} \sin kz + \Lambda(y) \nonumber \\
v(y,z) &=& V(y) \, \sqrt{2} \sin kz \\
w(y,z) &=& \frac1k \, V'(y) \, \sqrt{2} \cos kz  \ ,\nonumber
\end{eqnarray}
where the functions $U$, $V$, and $\Lambda$ satisfy the boundary
conditions
\begin{equation}  \label{eq:BC}
U'(a) = V(a) = V''(a) = \Lambda'(a) = 0
\ , \quad a=0,\,1 \ .
\end{equation}
Note that the vector field $\vectu$ defined in \eqref{eq:ansatz}
is automatically divergence-free.

This class of velocity fields $\vectu$ \eqref{eq:ansatz} is restrictive,
but
in our opinion it constitutes a physically reasonable ansatz.
It has been observed for plane parallel shear flows that the first modes
to
lose absolute stability have only cross-stream and span-wise variation with
no
dependence on the stream-wise coordinate $x$.
Moreover, the parameter $k$ in \eqref{eq:ansatz} can take any real value,
so
this does not impose any restriction on the wavelength of the pattern
in span-wise ($z$) direction.
Note also that the case of very high Reynolds numbers
corresponds to the choice $c=0$ (see \eqref{eq:L}),
and in this case the family \eqref{eq:ansatz}
will tend to the family \eqref{eq:u-k}
which we know achieves the upper bound
on the dissipation at infinite $Re$.
All these considerations make the choice of the family \eqref{eq:ansatz}
quite reasonable.
In the spirit of full disclosure, however, we reiterate emphatically 
the assumption that we make in the analysis that follows:

\medskip
\noindent
{\it Ansatz:} We assume that the maximizing vector fields for the
functional \eqref{eq:L} have the functional form \eqref{eq:ansatz}.

\bigskip

In terms of $U$, $V$, and $\Lambda$, the expression \eqref{eq:beta_c} for
the energy dissipation reads
\begin{eqnarray*}
\beta[U,V,\Lambda]
&=& (1+c) \frac{\langle \Phi'\Lambda\rangle
                \langle\Psi UV +\frac1{Re}\Psi'\Lambda\rangle}
        {\langle\Phi\Psi\rangle} \nonumber \\[2mm]
& & - \frac{c}{Re} \left\langle k^2U^2 + k^2V^2 + {U'}^2 + 2{V'}^2
                + \frac1{k^2}{V''}^2 + {\Lambda'}^2 \right\rangle \ ,
\end{eqnarray*}
and the functional $L[\vectu]$ \eqref{eq:L}
taking into account the normalization constraint becomes
\[
L[U,V,\Lambda] = \beta[U,V,\Lambda]
+ \frac{\lambda}{2} \left\langle U^2 + V^2
                + \frac1{k^2}{V'}^2 + {\Lambda}^2 -1\right\rangle
\ .
\]
The Euler-Lagrange equations for $U$, $V$, $\Lambda$ are
\numparts
\begin{eqnarray}
\fl
\frac{2c}{Re}\,U'' + \left(\lambda-\frac{2ck^2}{Re}\right) U
    + (1+c)\frac{\langle\Phi'\Lambda\rangle}{\langle\Phi\Psi\rangle}
        \, \Psi \, V &=& 0  \label{ELU} \\[4mm]
\fl
-\frac{2c}{Re \, k^2} \, V''''
    + \left(\frac{4c}{Re}-\frac{\lambda}{k^2}\right) V''
    + \left(\lambda - \frac{2ck^2}{Re}\right) V
    + (1+c)\frac{\langle\Phi'\Lambda\rangle}{\langle\Phi\Psi\rangle}
        \, \Psi \, U &=& 0  \label{ELV} \\[4mm]
\fl
\frac{2c}{Re}\, \Lambda'' + \lambda\Lambda
        + \frac{1}{Re}\,
        (1+c)\frac{\langle\Phi'\Lambda\rangle}{\langle\Phi\Psi\rangle}
                \, \Psi'
        + \left[(1+c)\frac{\langle\Psi UV\rangle}
                        {\langle\Phi\Psi\rangle}
        + \frac{1}{Re}(1+c)\frac{\langle\Psi'\Lambda\rangle}
                {\langle\Phi\Psi\rangle} \right] \Phi' &=& 0 \ ,
\label{ELL}  
\end{eqnarray}
\endnumparts
where the ``eigenvalue'' $\lambda$ is to be adjusted so that the triple
$(U,V,\Lambda)$ satisfies the normalization
\begin{equation}  \label{eq:norm}
\left\langle U^2 + V^2 + \frac1{k^2}{V'}^2
        + {\Lambda}^2 \right\rangle = 1 \ .
\end{equation}

%%%%%%%%%%%%%%%%%%%%%%%%%%%%%%%%%%%%%%%%%%%%%%%%%%%%%%%%%%%%

\subsection{Exact solution at finite $Re$ for the case $c=0$}

In the case $c=0$, the Euler-Lagrange equations
\eqref{ELU}, \eqref{ELV}, \eqref{ELL} become
\numparts
\begin{eqnarray}
\lambda \, U
        + \frac{\langle\Phi'\Lambda\rangle}{\langle\Phi\Psi\rangle}
        \, \Psi \, V &=& 0
                \label{c=0,genRe:1}  \\[2mm]
-\frac{\lambda}{k^2}\, V''
    + \lambda \, V
    + \frac{\langle\Phi'\Lambda\rangle}{\langle\Phi\Psi\rangle}
        \, \Psi \, U &=& 0
                \label{c=0,genRe:2}  \\[2mm]
\lambda \, \Lambda
        + \frac{1}{Re}\,
        \frac{\langle\Phi'\Lambda\rangle}{\langle\Phi\Psi\rangle}
                \, \Psi'
        + \left[\frac{\langle\Psi UV\rangle}
                        {\langle\Phi\Psi\rangle}
        + \frac{1}{Re} \, \frac{\langle\Psi'\Lambda\rangle}
                {\langle\Phi\Psi\rangle} \right] \Phi' &=& 0 \ .
                \label{c=0,genRe:3}
\end{eqnarray}
\endnumparts
Then the equations for $U$ and $\Lambda$ are algebraic equations, so the
only boundary conditions that have to be satisfied are
\begin{equation}  \label{eq:bc_U_c=0}
V(a) = 0 \quad \mbox{for } a=0, 1 \ .
\end{equation}

We can solve the boundary value problem \eqref{c=0,genRe:1},
\eqref{c=0,genRe:2}, \eqref{c=0,genRe:3}, \eqref{eq:bc_U_c=0} explicitly.
First, expressing $U$ from \eqref{c=0,genRe:1},
and substituting into \eqref{c=0,genRe:2},
we obtain the following boundary value problem for $V$:
\begin{equation}  \label{c=0,genRe:V}
-\frac{1}{k^2}\, V''
    + V
    = E^2 \, \frac{\Psi^2}{\langle\Psi^2\rangle} \, V
\ , \qquad V(0) = V(1) = 0  \ ,
\end{equation}
where we have set
\begin{equation}  \label{eq:E}
E := \frac{\langle\Phi'\Lambda\rangle \sqrt{\langle\Psi^2\rangle}}
                {\lambda \langle\Phi\Psi\rangle}  \ .
\end{equation}
For each choice of test function $\Psi$ we obtain a sequence of functions
$V_n$  and numbers $E_n$, $n=1,2,3,\ldots$.
For each $n$, the numbers $E_n$ and the functions
$V_n$ depend on $Re$, $k$, and the choice of test function $\Psi$.
The functions $\Lambda_n$ are (see the Appendix for a derivation)
\begin{equation}  \label{eq:lambda-exact}
\fl
\Lambda_n(y)
=
\left[
 - \frac{1}{\sqrt{3}} \,
  \sqrt{1
  + \frac{E_n^2}{Re^2 \langle\Psi^2\rangle}
        \left( \langle\Psi'^2 \rangle
                + \frac{\langle \Phi'\Psi'\rangle^2}{3} \right)
        }
+ \frac{E_n\langle\Phi'\Psi'\rangle}{3Re\sqrt{\langle\Psi^2\rangle}}
\right]
\Phi'(y)
- \frac{E_n}{Re\sqrt{\langle\Psi^2\rangle}}\,\Psi'(y) \ ,
\end{equation}
and the functions $U_n$ are
\begin{equation}  \label{eq:U-exact}
U_n(y) = -\frac{\Psi}{\sqrt{\langle\Psi^2\rangle}} \, V_n(y) \ .
\end{equation}
In the derivation of \eqref{eq:lambda-exact} we used the normalization
condition \eqref{eq:norm} so that it is automatically satisfied.
Then the (non-dimensional) energy dissipation rate is
\begin{eqnarray}   \label{beta_general}
\beta_n
&=&
\frac{\langle\Phi'\Psi'\rangle} {3\langle\Phi\Psi\rangle} \frac1{Re}
+
\frac{\langle\Phi'\Psi'\rangle}
        {3\langle\Phi\Psi\rangle \langle\Psi^2\rangle}
        \left( \langle\Psi'^2\rangle
                - \frac{\langle\Phi'\Psi'\rangle^2}{9} \right)
\frac{E_n^2}{Re^3}  \nonumber \\[2mm]
& & 
+\frac{\sqrt{\langle\Psi^2\rangle}} {3\sqrt{3}\langle\Phi\Psi\rangle}
\frac{1}{E_n}
\left[1+\frac{1}{\langle\Psi^2\rangle}
                \left(\langle\Psi'^2\rangle
                +       \frac{\langle\Phi'\Psi'\rangle^2}{3}\right)
                \frac{E_n^2}{Re^2}\right]^{3/2}
\ .
\end{eqnarray}
What remains to be done for a given shape potential and multiplier
function
is to find the solutions for $V$ and $E$.
This we do numerically.

%%%%%%%%%%%%%%%%%%%%%%%%%%%%%%%%%%%%%%%%%%%%%%%%%%%%%%%%%%%%

\subsection{Finding the velocity profile
        and energy dissipation for $c>0$}

Suppose that we have found the functions
$U^{(0)}_n$ \eqref{eq:U-exact},
$V^{(0)}_n$ \eqref{c=0,genRe:V},
and $\Lambda^{(0)}_n$ \eqref{eq:lambda-exact}
satisfying the Euler-Lagrange equations
\eqref{c=0,genRe:1}, \eqref{c=0,genRe:2}, \eqref{c=0,genRe:3}
and the boundary conditions \eqref{eq:bc_U_c=0}
in the case $c=0$.
In order to find the solution
$U_n$, $V_n$, $\Lambda_n$ of the boundary value problem
\eqref{ELU}, \eqref{ELV}, \eqref{ELL}, \eqref{eq:BC}
that satisfy the normalization condition \eqref{eq:norm}
for $c>0$,
we use Newton method with
$U^{(0)}_n$, $V^{(0)}_n$, $\Lambda^{(0)}_n$
as initial guess.

According to the general methodology of the mini-max procedure,
we have to first maximize the expression for the energy dissipation rate
$\beta$ over all allowed velocity fields $\vectu$ \eqref{eq:ansatz},
and then to minimize $\max_\vectu \beta$
over all allowed functions $\Psi$.
With our ansatz for the form of $\vectu$,
maximizing over $\vectu$ means maximizing over
all real values of $k$.
Then having found the maximum of $\beta$ over $k$,
we minimize over both $\Psi$ {\em and} the balance parameter $c\geq 0$.
In practice we have to choose a particular family of test functions $\Psi$
depending on a small number of parameters, and minimize over those
parameters and $c$.
We will take a 1-parameter family of test functions $\Psi_\delta$ (given
explicitly in \eqref{eq:Psi-new} below) where the parameter $\delta$ is a
measure of the thickness of a ``boundary layer''.

Let $\beta(Re,\delta,c,k)$ be the mini-max upper bound
for the turbulent energy dissipation
as a function of the Reynolds number $Re$,
the parameter $\delta$ of the family $\Psi_\delta$,
the balance parameter $c$, and the wavenumber~$k$.
Define $\beta^*(Re,\delta,c)$ to be the maximum over $k$
of $\beta(Re,\delta,c,k)$, and $k^*(Re,\delta,c)$ to be
the value of $k$ for which $\beta(Re,\delta,c,k)$
attains this maximum.
Then
\begin{equation}  \label{eq:beta*}
\fl
\beta^*(Re,\delta,c) 
:=
\max_k \beta(Re,\delta,c,k) \ , 
\qquad
k^*(Re,\delta,c) 
:= 
\mathrm{argmax}\,\,\beta(Re,\delta,c,\cdot) \ .  
\end{equation}
After maximizing over $k$, i.e., over the family
of velocity fields $\vectu$ \eqref{eq:ansatz},
we minimize over the parameter $\delta$ of the family of test
functions $\Psi_\delta$, and the balance parameter $c$.
That is, we compute
\begin{equation}  \label{eq:beta_b}
\fl
\beta_b(Re) := \min_{(\delta,c)} \beta^*(Re,\delta,c)
\ , \qquad
(\delta^*(Re),c^*(Re))
        := \mathrm{argmin}\,\,\beta^*(Re,\cdot,\cdot) \ .
\end{equation}

%%%%%%%%%%%%%%%%%%%%%%%%%%%%%%%%%%%%%%%%%%%%%%%%%%%%%%%%%%%%

\section{Numerical results}  \label{sec:num-results}

\subsection{Numerical example and implementation}

As a specific model to analyze we chose the same shape function
$\phi$ as in \cite{DoeringES03}:
\[
\Phi(y) = \frac{\sqrt{2}}{\pi} \sin{\pi y} \ , \quad
\phi(y) = -\Phi'(y) = - \sqrt{2} \cos{\pi y} \ .
\]
In \cite{DoeringES03}, the test functions $\Psi_\delta$ were chosen
piecewise linear but not continuously differentiable.
For computational reasons we replace them
with the smooth family
\begin{equation}  \label{eq:Psi-new}
\Psi_\delta(y)
=
(1-\E^{-y/\delta}) \, (1-\E^{-(1-y)/\delta})
\ , \quad \delta > 0 \ .
\end{equation}
The functions \eqref{eq:Psi-new} satisfy the boundary conditions
$\Psi_\delta(0) = 0 = \Psi_\delta(1)$.

The boundary conditions of the Euler-Lagrange equations naturally suggest
the use of Chebyshev polynomials as interpolants to implement a 
pseudo-spectral scheme \cite{Trefethen00} 
to solve these equations.  The Matlab differentiation matrix 
suite~\cite{WeidemanR00} 
simplifies the implementation by providing routines to
discretize and represent differentiation operators as matrices. 
Differentiation 
of a function then becomes multiplication of the differentiation matrix
with the vector of the function values at those Chebyshev nodes.  
However, the discretized
equations are still nonlinear in the $c \neq 0$ case.  
We started with the
$c = 0$ equations which are solvable as a linear eigenvalue  
problem~\eqref{c=0,genRe:V}.  
Then the standard Newton's method was applied 
to these solutions and iterated to solve
the nonlinear equations \eqref{ELU}, \eqref{ELV}, \eqref{ELL}.  
The Jacobian matrices needed in the Newton's method
were computed by a simple forward difference scheme.  Throughout all 
computations, 128 and 64 Chebyshev nodes were used 
(the differences between the results for these 
choices of number of nodes did not exceed~$10^{-7}$).  

\begin{figure}
\centering
\includegraphics[width=0.8\textwidth]
        {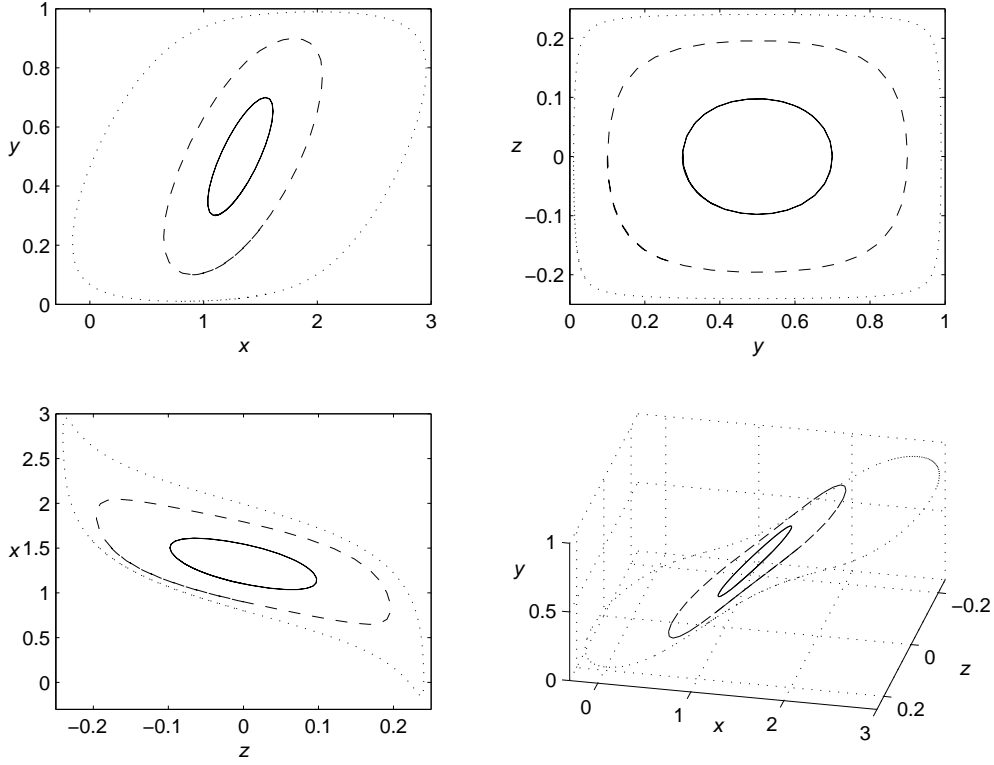}
\caption{Integral lines of the velocity field for $\re=50$.}
\label{fig:flow_50}
\end{figure}

\begin{figure}
\centering
\includegraphics[width=0.8\textwidth]
        {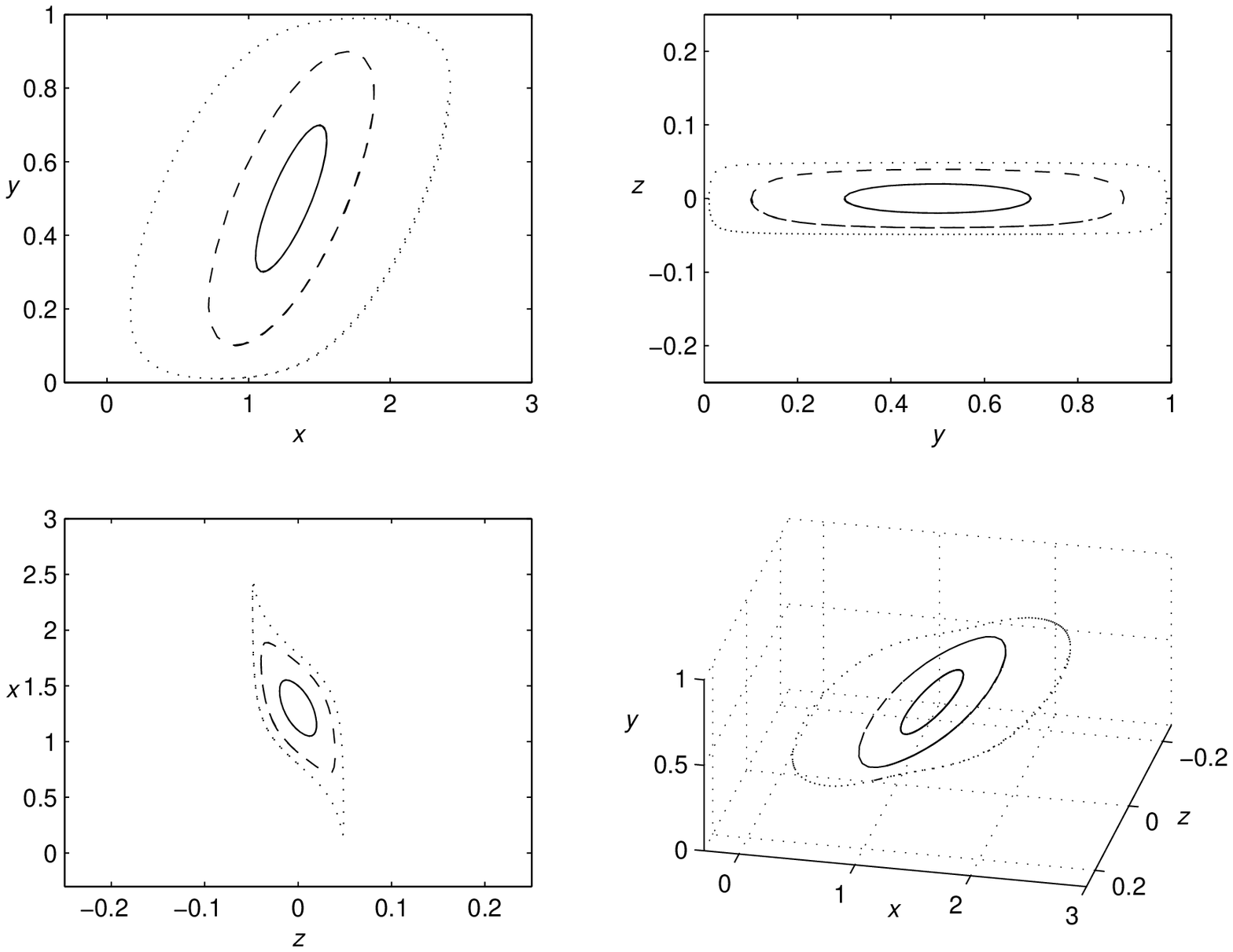}
\caption{Integral lines of the velocity field for $\re=1000$.}
\label{fig:flow_1000}
\end{figure}

To illustrate the typical geometry of the flow, 
in Figures~\ref{fig:flow_50} and \ref{fig:flow_1000}, 
we show the three coordinate projections 
and the 3-dimensional view of 
typical integral lines 
(i.e., solutions of $(\dot x,\dot y,\dot z)=(u,v,w)$ 
for $(u,v,w)$ given by \eqref{eq:ansatz})
of the maximizing flow field 
for $\re=50$ and $\re=1000$, respectively.  
The  values of the parameters $\delta$, $c$, $k$, 
for the fields shown are the ones that give the optimal bound, 
$\beta_b(\re)$ given by \eqref{eq:beta_b}.  

\begin{figure}
\centering
\includegraphics[width=0.8\textwidth]
        {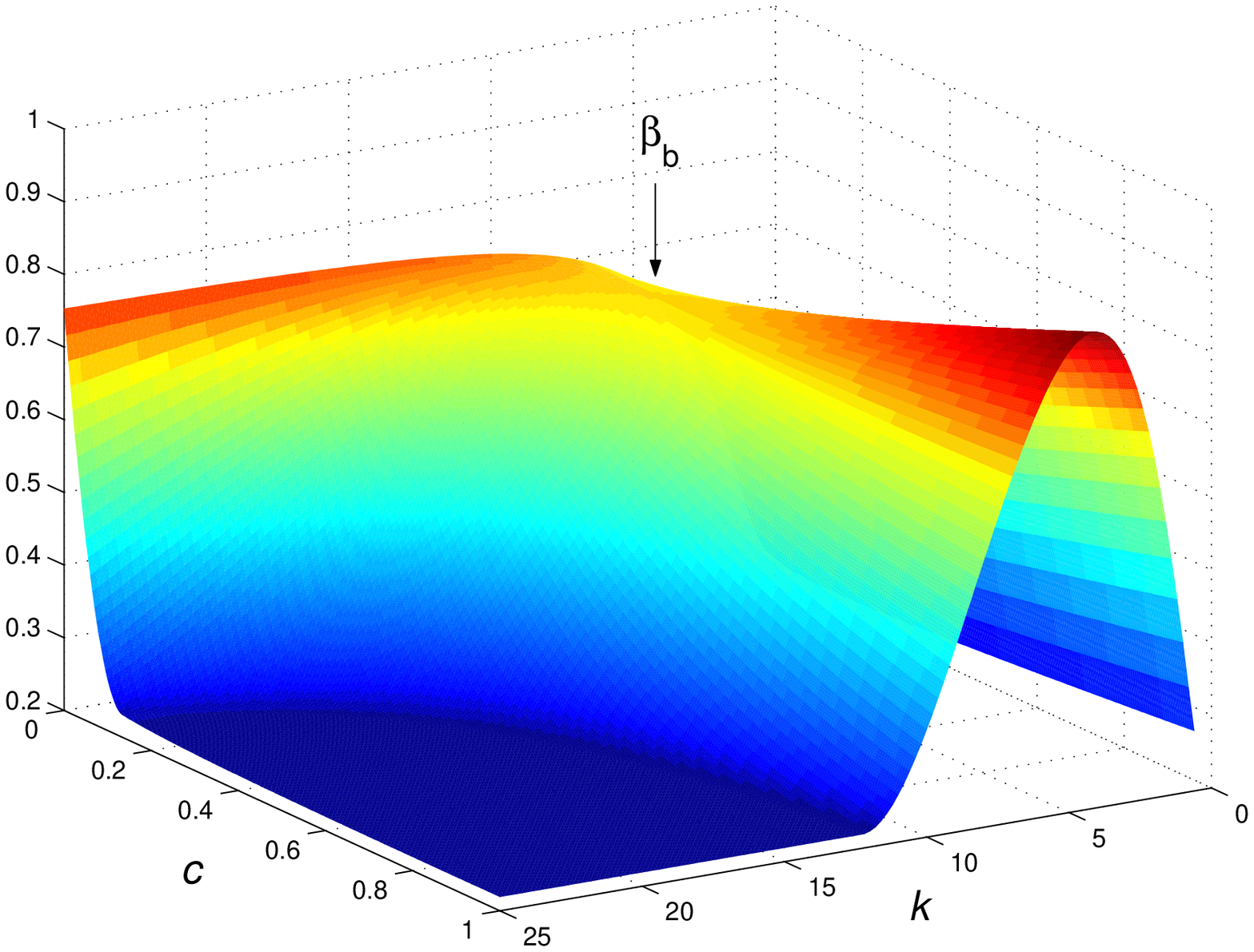}
%       {plot_exp_delta_0.040_c_0_0.01_0.5_k_1_15_Re_35.eps}
\caption{Bound on dissipation for $Re=50$ 
        as a function of $c$ and $k$
        (using $\Psi_{0.04}$).}
\label{fig:3d-plot}
\end{figure}

As an example of the mini-max procedure,  
we show in Figure~\ref{fig:3d-plot} 
%
%  To make the figure figure-3d-Re-50.eps, do:
%  load beta_ck_exp_delta_0.040_c_0_0.005_1_k_0.6_0.2_25_Re_50_3d_picture_v6.mat
%  (this file is in TURBULENCE/MATLABFILES/computeBeta/ on s0077
%  [R,C,K] = size(beta)
%  for cc=1:C, for kk=1:K b50(cc,kk) = beta(1,cc,kk); end, end
%  hold off; surf(k,c,b50), shading flat, colormap(jet)
the upper bound on the dissipation for $Re=50$
obtained by using as a test function
$\Psi_\delta$ from \eqref{eq:Psi-new} with $\delta=0.04$;
the bound is given as a function of $c\in[0,1]$ and $k\in(0,25]$.

%  THIS COMMENT IS OLD!!!!!!  THE RIGHT ONE IS ABOVE!!!!
%To make the figure plot_exp_delta_0.050_k_3_15_Re_75_vs_c.eps, do
%>> load beta_ck_exp_delta_0.050_c_0_0.01_0.5_k_1_15_Re_25_5_100.mat
%>> size(beta)
%ans =
%    16    51    15
%>> hold off; for i=1:15, plot(c(1:31),beta(11,1:31,i),'-'); axis([0 0.30
% 0.50 0.75]), hold on; end, hold off;
%and then put the labels k=3, ...

\begin{figure}
\centering \includegraphics[width=0.8\textwidth]
        {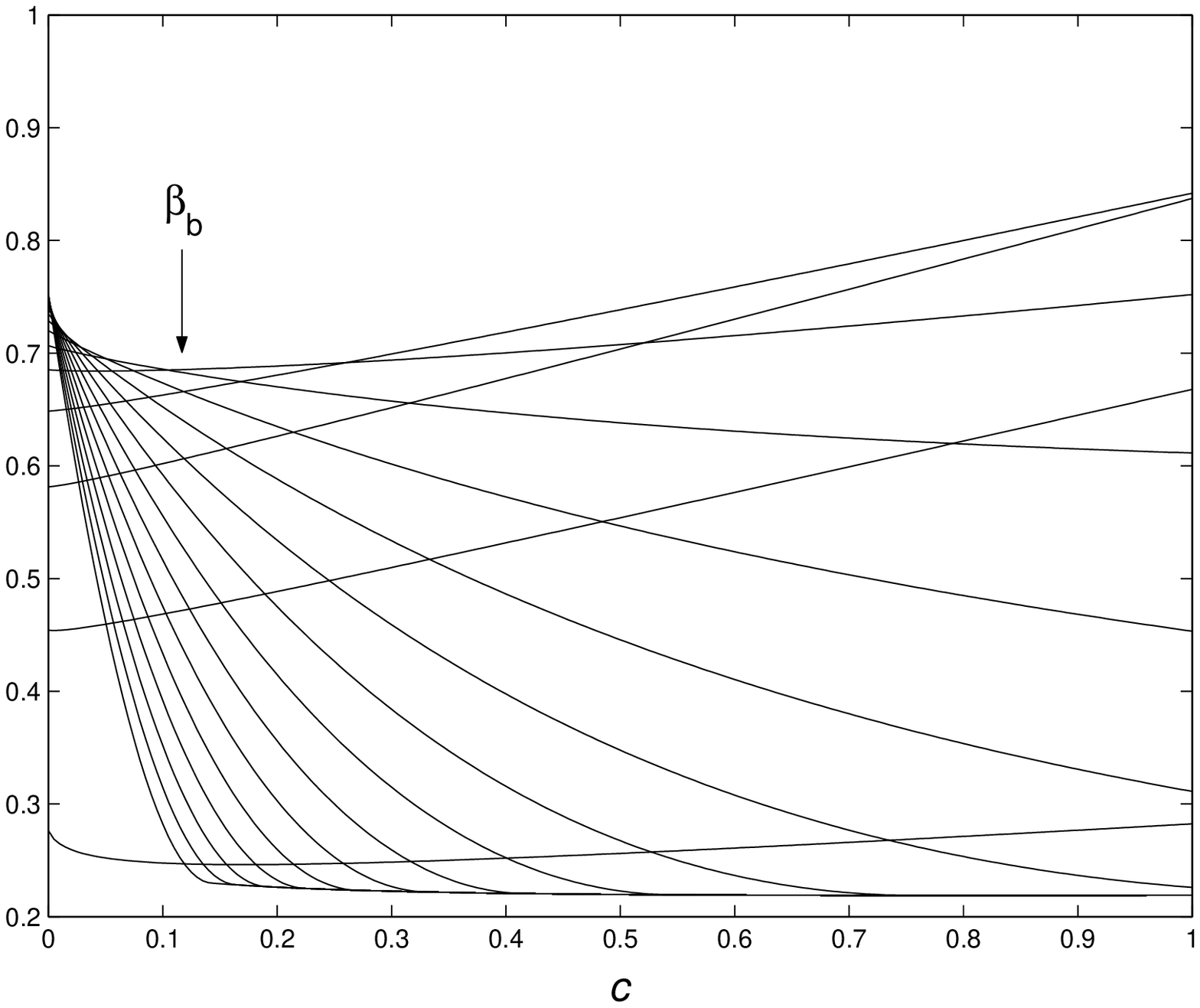}
%       {plot_exp_delta_0.050_k_3_15_Re_75_vs_c.eps} 
\caption{Bound on $\beta$ for $Re=50$ (obtained with $\Psi_{0.04}$)
        as a function of $c$ for several values of~$k$.}
\label{fig:beta-vs-c-many-k} 
\end{figure}

In Figure~\ref{fig:beta-vs-c-many-k} we show the bound on the dissipation
$\beta$ for $Re=50$ as a function of the balance parameter $c$ for
different values of the span-wise wavenumber $k$; the data presented have
been obtained with $\Psi_\delta$ with $\delta=0.04$. 
%  To make the figure beta-vs-c-many-k.eps, do 
%  load beta_ck_exp_delta_0.040_c_0_0.005_1_k_0.6_0.2_25_Re_50_3d_picture_v6.mat
%  (this file is in TURBULENCE/MATLABFILES/computeBeta/ on s0077
%  [R,C,K] = size(beta)
%  hold off; for i=1:7:K, plot(c(1:C),beta(1,1:C,i),'-'); hold on; end, hold off;
The figure illustrates the 
general behavior of $\beta$ as a function of $k$ and $c$ -- namely, for
small $k$, the value of $\beta$ increases with $c$, while for larger $k$, $\beta$
decreases with $c$.  
Clearly, the family of lines in the figure has an envelope 
-- this envelope is the graph of the function 
$\beta^*(50,0.04,c)$ \eqref{eq:beta*}.  
Having obtained the envelope, 
we find the minimum value of $\beta^*(50,0.04,c)$ 
-- this is the mini-max value we are looking for; 
this point is labeled with $\beta_b$ 
in Figures \ref{fig:3d-plot} (where it is the saddle point) 
and~\ref{fig:beta-vs-c-many-k}.

\subsection{Results}

\begin{figure} 
\centering
\includegraphics[width=0.8\textwidth]{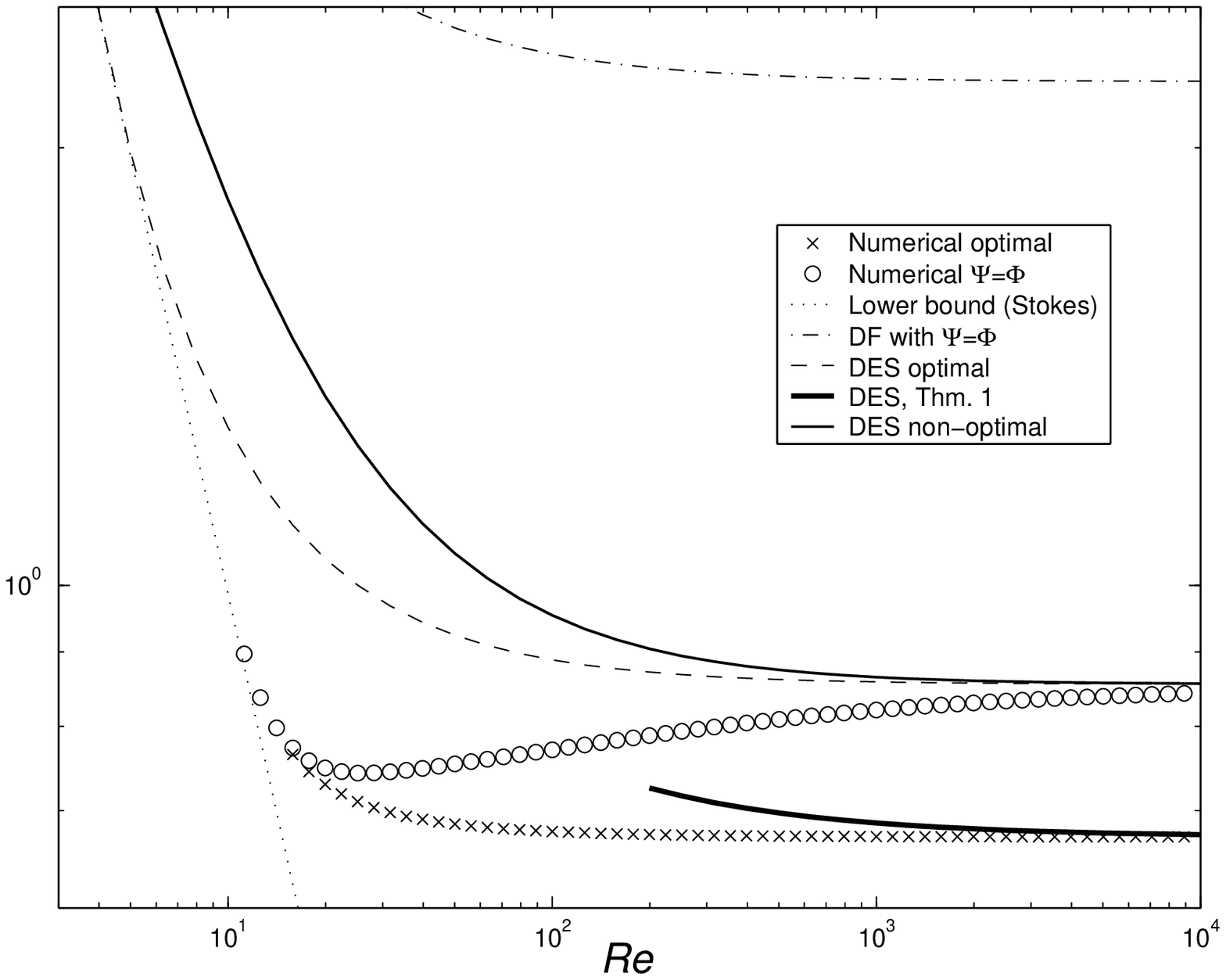} %{data_bounds2.eps}
\caption{Upper and lower bounds on $\beta$.} 
\label{fig:bounds}
\end{figure}

In Figure~\ref{fig:bounds}, 
we present the bounds from previous papers, as well as our
new numerical results.  The dotted straight line represents the lower
limit on the dissipation corresponding to Stokes (laminar) flow, 
\[
\beta_{\mathrm{Stokes}} \geq \frac{\pi^2}{Re} \ . \] 
The dot-dashed line
in the upper part of the figure is the bound following \cite{DoeringF02}
for this problem obtained with $\Psi = \Phi$:  
\[ \beta_{\mathrm{DF}} \leq
        \frac{\pi}{\sqrt{2}} + \frac{\pi^2}{Re} \ . 
\] 
The thin solid line
shows the ``non-optimal'' bound from \cite{DoeringES03} (equation (3.14)
in \cite{DoeringES03}), 
\[ \beta_{\mathrm{DES,\ non-optimal}} \leq
\frac{\sqrt{2}\pi}{\sqrt{27}} + \frac{\pi^2}{Re} \ , \] 
while the
long-dashed one gives their ``optimal'' estimate (obtained from equation
(3.12) in \cite{DoeringES03} by first minimizing over $\xi$ and then
plugging $\Psi=\Phi$): 
\[ \beta_{\mathrm{DES,\ optimal}} \leq
\frac{\sqrt{2}\pi}{\sqrt{27}}
                \left(1+\frac{2\pi^2}{3Re^2}\right)^{3/2}
        + \frac{\pi^2}{3Re}
                \left(1+\frac{4\pi^2}{9Re^2}\right) \ .
\]
(Note that this line bifurcates from the lower Stokes bound at
$Re=\sqrt{2}\pi\approx 4.4429$).
The thick solid line starting from $Re \approx 178$
is the best upper bound for high values of $Re$
from Theorem~1 of \cite{DoeringES03}:
\begin{equation}  \label{eq:DES-thm1}
\beta_{\mathrm{DES,\ Thm.\ 1}}
\leq \frac{\pi^2}{\sqrt{216}}
+ \frac{5(6\pi^2)^{1/5}}{4^{4/5}\,Re^{4/5}}
\approx
0.67154 + \frac{3.73089}{Re^{4/5}} \ .
\end{equation}
The circles in the figure give our new numerically determined upper
bounds  on $\beta$ with the choice $\Psi=\Phi$ and the crosses represent
our
numerical results for the choice \eqref{eq:Psi-new} of $\Psi_\delta$.

\begin{figure} 
\centering
\includegraphics[width=0.8\textwidth]{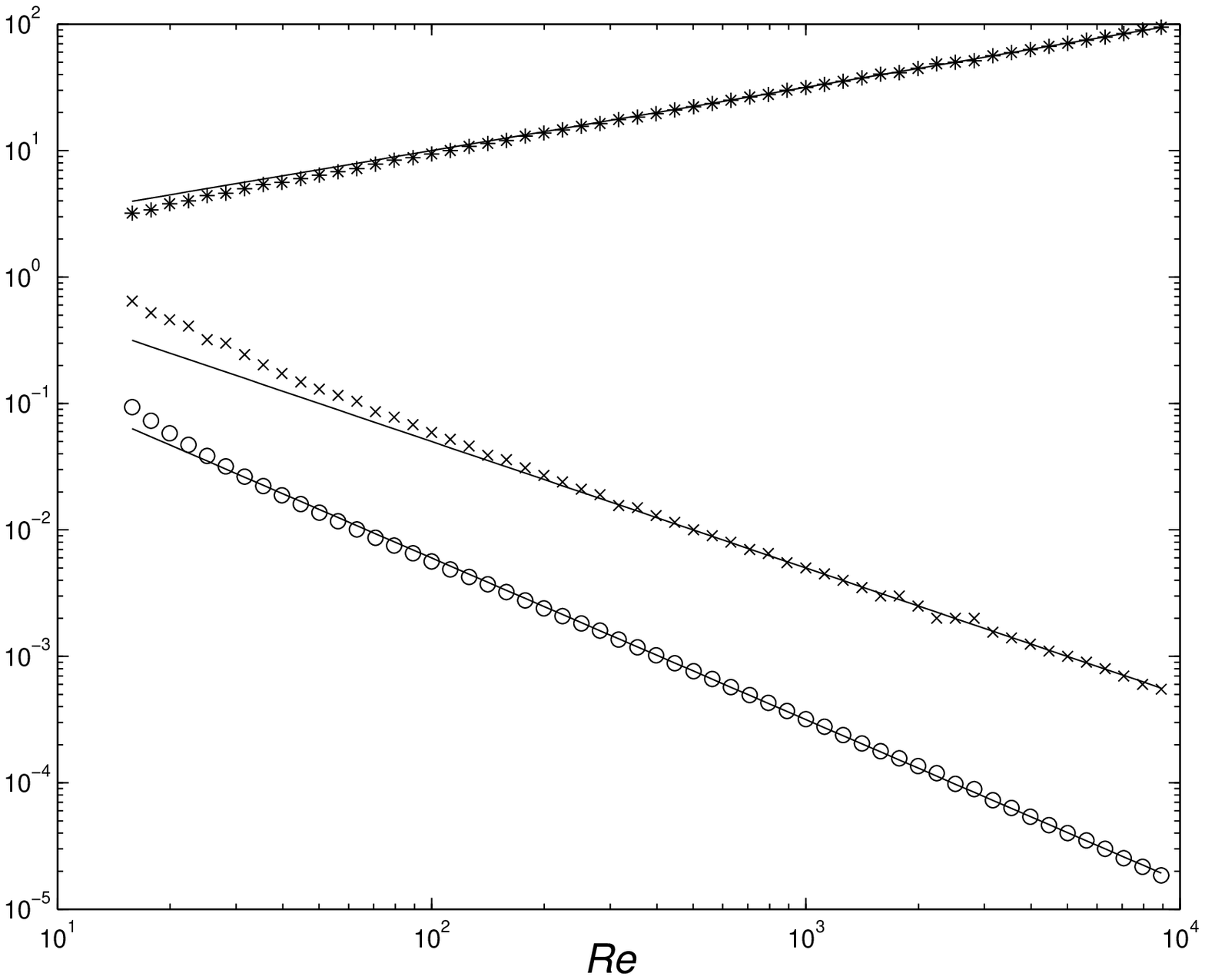} 
                        %{log_log_high_re.eps}
\caption{Power-law behavior of $\beta-\frac{\pi^2}{\sqrt{216}}$, 
$k^*$ and $c^*$ vs.~$\re$ (see the text).} 
\label{fig:log-log} 
\end{figure}

In Figure~\ref{fig:log-log} we have plotted 
$\beta-\frac{\pi^2}{\sqrt{216}}$ (circles), 
$k^*$ (stars), and $c^*$ (x's), 
versus $Re$ for the values of the
dissipation bound obtained using the function $\Psi_\delta$ from
\eqref{eq:Psi-new}.  
We see that $k^*\sim \sqrt{\re}$, $c^*\sim\frac{1}{\re}$, 
and from the figure 
we observe that $\beta-\frac{\pi^2}{\sqrt{216}}$ 
also behaves like a power of~$\re$.  
In the figure we illustrate these behaviors 
by showing the straight lines 
\[
\beta = \frac{\pi^2}{\sqrt{216}} + \frac{2.158}{\re^{1.28}} \ , 
\quad 
k^* = 1.0\,\sqrt{\re} \ , 
\quad 
c^* = \frac{5.0}{\re} \ .
\]
%
%and linear regression on the points with $Re = 55$ through 3000
%indicates that 
%\[ 
%\log_{10}\left(\beta-\frac{\pi^2}{\sqrt{216}}\right) =
%a\log_{10}Re + b
%\] 
%with $a= -1.213\pm0.005$, $b = 0.168\pm0.012$ (the
%error is only the error of the regression). This corresponds to the
%behavior 
%\[ 
%\beta \approx \frac{\pi^2}{\sqrt{216}} + \frac{1.4}{Re^{1.2}}
%\] 
%which is to be compared with equation \eqref{eq:DES-thm1}.  Not
%unexpectedly, the family $\Psi_\delta$ from \eqref{eq:Psi-new} tends
%toward the sharp infinite $Re$ bound \eqref{eq:DES-thm1} derived in
%\cite{DoeringES03}.  
%
% regression with the point re = 50 
%Regression coefficient (SLOPE)  = -1.213214 
%Standard error of coefficient = 0.004403908 
%t - value for coefficient = -275.4858 
%Regression constant (INTERCEPT)  = 0.1664848 
%Standard error of constant = 0.01066843 
%regression starting at re = 55 
%Regression coefficient (SLOPE)  = -1.214162
%Standard error of coefficient           = 0.004613178
%t - value for coefficient               = -263.1942
%Regression constant (INTERCEPT)                 = 0.169191
%Standard error of constant              = 0.01132384

%%%%%%%%%%%%%%%%%%%%%%%%%%%%%%%%%%%%%%%%%%%%%%%%%%%%%%%%%%%%
%%%%%%%%%%%%%%%%%%%%%%%%%%%%%%%%%%%%%%%%%%%%%%%%%%%%%%%%%%%%

\section{Concluding remarks}  \label{sec:concl}

We have derived new bounds on the energy dissipation rate for an example
of body-force driven flow in a slippery channel.
The fundamental improvement over previous results came from the
application of the balance parameter in the variational formulation of the
bounds, together with numerical solution of the Euler-Lagrange equations
for the best estimate.

\begin{figure} 
\centering
\includegraphics[width=0.8\textwidth]{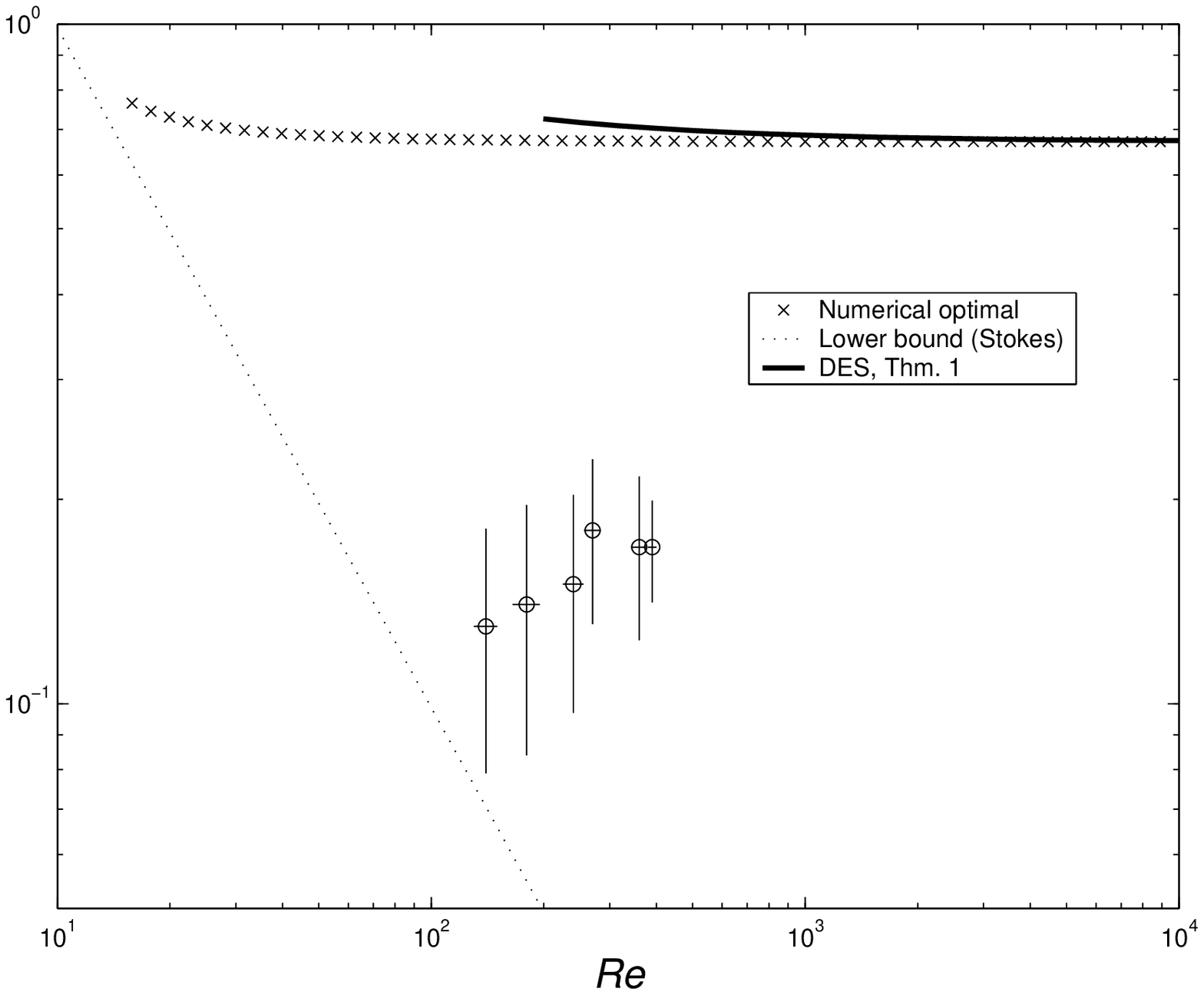} 
\caption{Comparison between theoretical results 
and DNS data (same symbols as in Figure~\ref{fig:bounds}).} 
\label{fig:theor-exp} 
\end{figure}

In Figure \ref{fig:theor-exp} the results 
of this analysis are compared with the direct
numerical simulations of the three-dimensional Navier-Stokes equations
first reported in \cite{DoeringES03}.  
Over the Reynolds number range 100--1000 where the data lie, the best
bounds derived here, using the balance parameter and minimization over the
(restricted) family of multiplier functions $\Psi_{\delta}$, result in a
quantitative improvement over the previous rigorous estimates.
We observe that the measured dissipation is a factor of 3 to 4 below the
bound, which should be considered nontrivial given the {\it a priori}
nature of the estimates derived here.
Presumably a full optimization over possible multiplier functions $\Psi$
would result in a further lowering of the estimate at lower values of
$Re$, producing a bound that intersects the lower Stokes bound right at
the energy stability limit (which we compute to be at $Re = 2 \pi$).
We note from Figure \ref{fig:bounds} 
that the bounds computed with $\Phi_{\delta}$ tend
to agree with those computed using $\Phi = \Psi$ at lower values of $Re$,
indicating that both trial functions are about the same ``distance'' from
the true optimal multiplier.

At higher Reynolds numbers the optimal solutions computed here converge
rapidly to the asymptotic bound $\beta_b(\infty)$ computed analytically
in \cite{DoeringF02}.  
Indeed, the bound derived here approaches the asymptotic limit with a
difference vanishing $\sim Re^{-1.28}$.
This particular scaling of the approach to the asymptotic limit helps to
understand the role that the balance parameter plays to lower the bound:
while a naive estimate suggests that the approach might be ${\cal
O}(Re^{-1})$, the faster convergence may be attributed to the interplay of
the $c \sim Re^{-1}$ and $k \sim \sqrt{Re}$ scaling in the prefactor and
the subtracted term in~\eqref{eq:beta_c}.  

There are several directions in which this line of research could be
continued.
One is to develop more reliable and accurate analytical methods for
estimating the best bounds at finite $Re$.
This would probably involve asymptotic approximations for small but finite
values of $Re^{-1}$ which could lead to more general applications for
other variational problems as well.
Another direction would be to develop methods to determine the true
optimal multiplier function at finite $Re$.
The motivation there would largely be as a point of principle, to demonstrate
that the full min-max procedure can indeed be carried out---at least for
simple set-ups such as those considered here.
Finally, going beyond the simple $\sin{\pi y}$ forcing considered in this
paper there remains the question, first posed in \cite{DoeringES03}, 
as to the connection between the optimal multiplier
and the true mean profile realized in direct numerical simulations.
Specifically, the question is whether there is a sensible correspondence
between the shape of the optimal multiplier and the mean profile for
general force shapes.
The idea is that the optimal multiplier contains information about the
extreme fluctuations that might be realized in a turbulent flow, and some
of those features may correlate with the statistical properties of the
flows.

%%%%%%%%%%%%%%%%%%%%%%%%%%%%%%%%%%%%%%%%%%%%%%%%%%%%%%%%%%%%
%%%%%%%%%%%%%%%%%%%%%%%%%%%%%%%%%%%%%%%%%%%%%%%%%%%%%%%%%%%%

\ack

This work was supported in part by National
Science Foundation Award PHY-0244859.  
The numerical computations were performed 
in the Department of Mathematics at the University of Texas at
Austin.

%%%%%%%%%%%%%%%%%%%%%%%%%%%%%%%%%%%%%%%%%%%%%%%%%%%%%%%%%%%%
%%%%%%%%%%%%%%%%%%%%%%%%%%%%%%%%%%%%%%%%%%%%%%%%%%%%%%%%%%%%
%%%%%%%%%%%%%%%%%%%%%%%%%%%%%%%%%%%%%%%%%%%%%%%%%%%%%%%%%%%%

\appendix
\section*{Appendix: Derivation of the expression
        \eqref{eq:lambda-exact} for $\Lambda$}
\setcounter{section}{1}

In this Appendix we show how to derive the expression
\eqref{eq:lambda-exact} for $\Lambda$ in the case $c=0$. First exclude $U$
from \eqref{c=0,genRe:3} with the help of \eqref{c=0,genRe:1}:
\begin{equation} \label{c=0,genRe:L} 
\Lambda =
\frac{\langle\Phi'\Lambda\rangle \langle \Psi^2V^2\rangle}
               {\lambda^2 \langle\Phi\Psi\rangle^2} \, \Phi'
        - \frac{1}{Re}\, \frac{1}{\lambda \langle\Phi\Psi\rangle} \,
                \Bigl( \langle\Phi'\Lambda\rangle \, \Psi'
                       + \langle\Psi'\Lambda\rangle \, \Phi'
                \Bigr)  \ . 
\end{equation} 
%Now we will use \eqref{c=0,genRe:V} and \eqref{c=0,genRe:L} 
%together with the normalization \eqref{eq:norm} 
%to obtain the explicit expression for $\Lambda$ 
%\eqref{eq:lambda-exact}. 
Now multiply the equation for $U$
\eqref{c=0,genRe:1} by $-U$, add it to the equation for $V$
\eqref{c=0,genRe:2} multiplied by $V$, and integrate the resulting
identity to get the {\em equidistribution property} $\langle U^2 \rangle =
\left\langle V^2 + \frac1{k^2} V'^2 \right\rangle$, so that the
normalization condition \eqref{eq:norm} can now be written as
\begin{equation} \label{eq:norm1} 
2 \left\langle V^2 + \frac1{k^2} V'^2
\right\rangle
        + \langle \Lambda^2\rangle = 1 \ . 
\end{equation} 
Multiplying
\eqref{c=0,genRe:V} by $V$ and integrating using the boundary conditions
\eqref{eq:bc_U_c=0}, we obtain 
\[ \left\langle V^2 + \frac1{k^2} V'^2
\right\rangle = \frac{E^2 \langle\Psi^2 V^2\rangle}{\langle\Psi^2\rangle}
\ , 
\] 
which, together with the new normalization \eqref{eq:norm1}, yields
\[ 
\frac{1-\langle\Lambda^2\rangle}{2} = \frac{E^2 \langle\Psi^2
V^2\rangle}{\langle\Psi^2\rangle} \ . 
\] 
This expression and the
definition of $E$ \eqref{eq:E} allow us to write the coefficient of the
term of order $Re^0$ in the right-hand side of \eqref{c=0,genRe:L} as 
\[
\frac{\langle\Phi'\Lambda\rangle \langle \Psi^2V^2\rangle}
               {\lambda^2 \langle\Phi\Psi\rangle^2} = \frac{E^2
\langle\Psi^2 V^2\rangle}
        {\langle\Psi^2\rangle \langle\Phi'\Lambda\rangle} =
\frac{1-\langle\Lambda^2\rangle}
        {2 \langle\Phi'\Lambda\rangle} \ . 
\] 
Using the above relationship
and expressing the Lagrange multiplier $\lambda$ from \eqref{eq:E}, we can
rewrite \eqref{c=0,genRe:L} as 
\begin{equation} \label{eq:Lambda1} 
\Lambda
= 
%&=& 
%\frac{1-\langle\Lambda^2\rangle} 
% {2 \langle\Phi'\Lambda\rangle} \, \Phi' 
%- \frac{1}{Re} \, 
% \frac{E}{\sqrt{\langle\Psi^2\rangle}} \, 
%\left( \frac{\langle\Psi'\Lambda\rangle} 
% {\langle\Phi'\Lambda\rangle} \, \Phi' 
% + \Psi' 
% \right)  \nonumber \\[2mm] 
%&=& 
\left(
\frac{1-\langle\Lambda^2\rangle}
        {2 \langle\Phi'\Lambda\rangle}
        - \frac{1}{Re} \,
        \frac{E}{\sqrt{\langle\Psi^2\rangle}} \,
        \frac{\langle\Psi'\Lambda\rangle}
                {\langle\Phi'\Lambda\rangle} \right) \, \Phi' -
\frac{1}{Re} \, \frac{E}{\sqrt{\langle\Psi^2\rangle}} \, \Psi' \ .
\end{equation} 
%which we temporarily denote as %\[ %\Lambda := \mu \Phi'
%- \frac{1}{Re}\,\frac{E}{\sqrt{\langle\Psi^2\rangle}}\,\Psi' \ . 
%\] %The unknown function $\Lambda$ 
%can be expressed from \eqref{eq:Lambda1} exactly. 
Let $\mu$ be the coefficient of $\Phi'$ in \eqref{eq:Lambda1},
i.e., 
\begin{equation} \label{eq:Lambda-mu} 
\Lambda := \mu \Phi' -
\frac{1}{Re}\,\frac{E}{\sqrt{\langle\Psi^2\rangle}}\,\Psi' \ .
\end{equation} 
From this expression we easily obtain (recall that
$\langle\Phi'^2\rangle=1$) 
\begin{eqnarray*} 
\langle\Lambda^2\rangle 
&=&
\mu^2 - \frac{2E\langle\Phi'\Psi'\rangle}
                {Re\sqrt{\langle\Psi^2\rangle}} \, \mu +
\frac{E^2\langle\Psi'^2\rangle}{Re^2\langle\Psi^2\rangle} \\[2mm] \langle
\Phi'\Lambda \rangle &=& \mu - \frac{E\langle\Phi'\Psi'\rangle}
                {Re\sqrt{\langle\Psi^2\rangle}} \\[2mm] \langle
\Psi'\Lambda \rangle &=& \langle\Phi'\Psi'\rangle \, \mu -
\frac{E\langle\Psi'^2\rangle}{Re\sqrt{\langle\Psi^2\rangle}} \ .
\end{eqnarray*} 
Plugging these expressions in the definition of the
coefficient $\mu$, 
\[ \mu = \frac{1-\langle\Lambda^2\rangle}
        {2 \langle\Phi'\Lambda\rangle} - \frac{1}{Re} \,
        \frac{E}{\sqrt{\langle\Psi^2\rangle}} \,
         \frac{\langle\Psi'\Lambda\rangle}
                {\langle\Phi'\Lambda\rangle} \ , 
\] 
we obtain the
following quadratic equation for $\mu$: 
\[ 
3 \mu^2
-\frac{2E\langle\Phi'\Psi'\rangle}{Re\sqrt{\langle\Psi^2\rangle}}\,\mu -
\left(
        1 + \frac{E^2\langle\Psi'^2\rangle}{Re^2\langle\Psi^2\rangle}
  \right) = 0 \ . 
\] 
The ``physical'' solution of this equation (the one
that has the right behavior in the limit $Re\to\infty$) is 
\[
 \mu = -
\frac{1}{\sqrt{3}} \,
  \sqrt{1
  + \frac{E^2}{Re^2 \langle\Psi^2\rangle}
        \left( \langle\Psi'^2 \rangle
                + \frac{\langle \Phi'\Psi'\rangle^2}{3} \right)
        } +
\frac{E\langle\Phi'\Psi'\rangle}{3Re\sqrt{\langle\Psi^2\rangle}} \ . 
\]
Plugging this into \eqref{eq:Lambda-mu}, we obtain the desired expression
\eqref{eq:lambda-exact}.

\section*{References}
\bibliographystyle{plain} % plain, alpha, abbrv, unsrt 
\bibliography{bal_par} 

\end{document}